\documentclass{IEEEtran}
\usepackage{graphicx}
\usepackage{amsmath}
\usepackage{amssymb}
\usepackage[T1]{fontenc}
\usepackage{lipsum}
\usepackage{comment}
\usepackage{float}
\usepackage{xcolor}
\usepackage{tikz}
\newcommand*\circled[1]{%
	\protect\tikz[baseline=(char.base)]{
		\scriptsize
		\protect\node[shape=circle,draw,inner sep=0.5pt] (char) {#1};
	}
}
\usepackage{enumerate,letltxmacro}
\LetLtxMacro\itemold\item
\renewcommand{\item}{\itemindent0.02in\itemold}
\pdfoutput=1

\begin{document}
\title{Cascading Power 
Outages Propagate 
Locally in an Influence Graph that is not the Actual Grid Topology}
\author{Paul~D.~H.~Hines, \IEEEmembership{Senior Member, IEEE}, 
		Ian~Dobson, \IEEEmembership{Fellow, IEEE}, 
		Pooya~Rezaei, \IEEEmembership{Student Member, IEEE}%
\thanks{%
P.H. was funded in part by US NSF Award Nos.~ECCS-1254549 and DGE-1144388, 
and US DTRA Award No.~HDTRA110-1-0088. I.D. was funded in part by NSF grant 
CPS-1135825.

P. Hines and P. Rezaei are with the School of Engineering and the Complex Systems Center, 
University of Vermont, Burlington, VT  05405, USA; e-mail: paul.hines@uvm.edu, prezaei@uvm.edu.

I. Dobson is with the ECpE department, Iowa State University, Ames IA 50011 USA; 
email: dobson@iastate.edu.

}}
\maketitle

\begin{abstract}
In a cascading power transmission outage, component outages propagate non-locally; after one component outages, the next failure may be very distant, both topologically and geographically. As a result, simple models of topological contagion do not accurately represent the propagation of cascades in power systems. However, cascading power outages do follow patterns, some of which are useful in understanding and reducing blackout risk. This paper describes a method by which the data from many cascading failure simulations can be transformed into a graph-based model of influences that provides actionable information about the many ways that cascades propagate in a particular system. The resulting ``influence graph'' model is Markovian, in that component outage probabilities depend only on the outages that occurred in the prior generation. To validate the model we compare the distribution of cascade sizes resulting from $n-2$ contingencies in a $2896$ branch test case to cascade sizes in the influence graph. The two distributions are remarkably similar. In addition, we derive an equation with which one can quickly identify modifications to the proposed system that will substantially reduce cascade propagation. With this equation one can quickly identify critical components that can be improved to substantially reduce the risk of large cascading blackouts.
\end{abstract}


\section{Introduction}

\IEEEPARstart{P}{ower}
systems are generally robust to small disturbances, 
but unexpected combinations of failures sometimes initiate long chains of cascading outages, which can result in massive and costly blackouts.
Because of the low-probability high-impact nature of cascading failures, there is limited empirical data from which to understand the many ways that cascades propagate through a power system.
Simulations of cascading mechanisms can produce data, but there has been insufficient progress in extracting insights and useful statistical information from these data.

One approach to obtaining useful information is to try and leverage 
successes from network science,
which has an extensive literature on cascading failure and (more generally) contagion.
Models of topological contagion 
in which failures spread locally from a node to its immediate neighbors~\cite{Watts:2002,Dodds:2005,Brummitt:2012} 
have provided insight into a variety of problems, such as disease propagation~\cite{hebert2010propagation,miller2014epidemic}.
Similar topological contagion models have been suggested as a tool for power systems analysis 
(e.g.,~\cite{Crucitti:2004,Kinney:2005}).

\begin{figure}[H]
	\includegraphics[width=1.0\columnwidth]{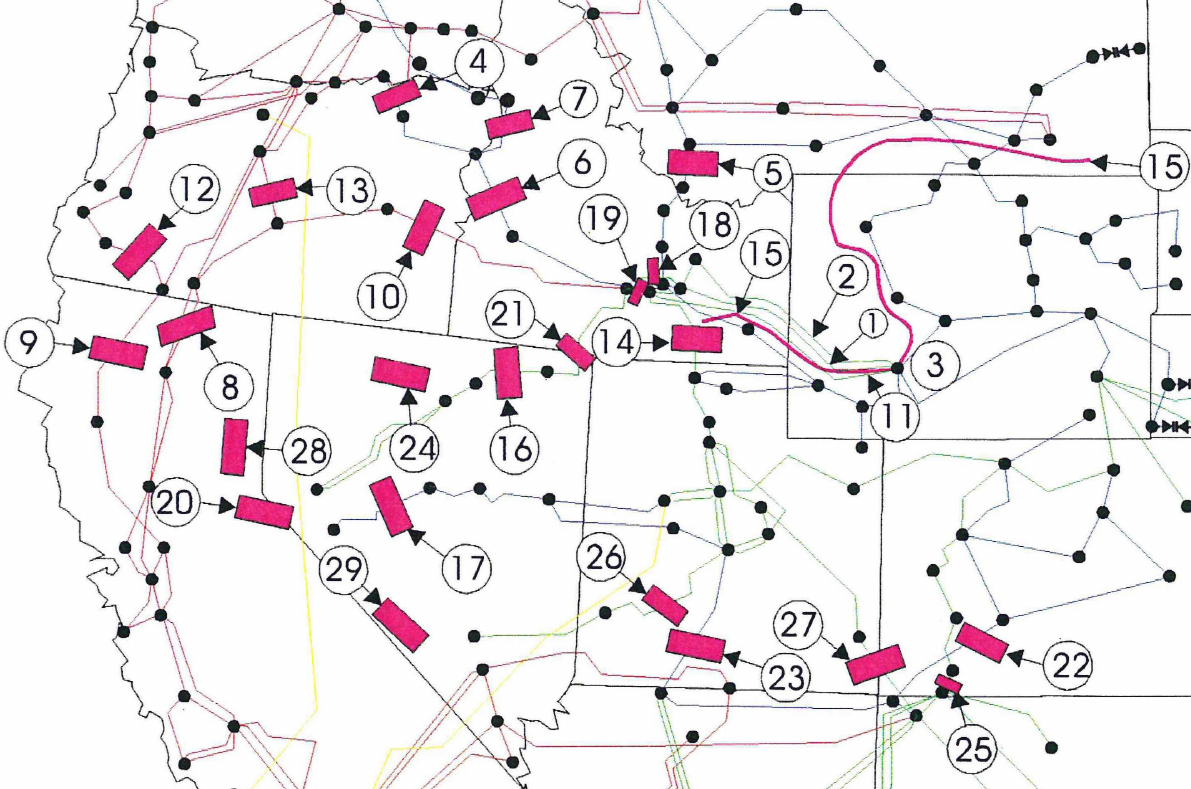}
	\caption{An illustration of the event sequence for the Western US blackout on July 2, 1996 (from~\cite{WSCC:1996}). 
	The sequence jumps across hundreds of kilometers at several points, 
	such as from \circled{3}- \circled{4} 
	and from \circled{7}- \circled{8}.
	}
	\label{wscc}
\end{figure}

This approach has two problems.
First, in power grids it is more appropriate to focus first on line (edge) outages, since line outages are an order of magnitude more likely than node (substation) outages. 
More importantly, as is well known to power systems engineers, cascading outages in power systems propagate non-locally as well as locally.
In a power grid, the next component to fail after a particular line outages may 
be very distant, both geographically and topologically~\cite{Dobson:2015}.
This can be seen very clearly from the sequence of events in the Western US blackout of 1996, shown in Fig.~\ref{wscc}, in which the failure sequence jumps across long distances at several points in the cascade.
Cascades in power networks are more similar to the 
random fuse model~\cite{de1985random} from statistical physics, in which non-local propagation does occur.

On the other hand, simulation models that capture aspects of detailed power grid physics and engineering are useful for understanding how cascades propagate.
Importantly these models do show the non-local propagation that is apparent in historical cascades.
Examples of physics-based models of cascading include
quasi-steady state models, such as DCSIMSEP~\cite{Eppstein:2012,Rezaei:2015rc}, OPA~\cite{Carreras:2002,Carreras:2013}, Manchester model~\cite{Nedic:2006}, and TRELSS~\cite{Bhavaraju:1992},
and dynamic models, such as COSMIC~\cite{Song:2015} and the hybrid model in~\cite{Henneaux:2015}.
While engineering models are useful, the data that result from this type of simulation are complicated and difficult to summarize, even for a single cascading failure simulation. 
In order to obtain useful statistical information, thousands, or even millions of simulations are often required, resulting in enormous sets of complicated output data.
To summarize and understand big data from detailed simulations, there is a need for higher-level models of cascading in large-scale systems that are simple enough to provide statistical insight, without abstracting away physical details in a way that could lead to erroneous conclusions.
Only a few such statistical models, which do not neglect the non-local nature of cascading, have been proposed in the literature.
Reference~\cite{CarrerasHICSS12} finds critical clusters of lines in simulated cascade data using a synchronization matrix, which determines the critical clusters as sets of lines that frequently overload in the same cascade and in a cascade that leads to a large blackout. 
This approach does not consider the order in which the lines overload during the cascade, but does indicate combinations of critical lines that are associated with blackouts.
The general idea of building an influence graph describing ways that component failures might spread goes back to~\cite{AsavathirathamCSM01,RoyHICSS01}.

This form of influence graph couples Markov processes at each node, but 
it remains unknown how to construct these influence graphs from data.
Stochastic models in~\cite{Wang:2012,Rahnamay-Naeini:2014} use Markov-chains to represent key characteristics of simulated cascades.
In prior work, the authors proposed a ``line-interaction graph'' approach to modeling cascades from data~\cite{Hines:2013hicss}.
A matrix-based approach, which is in some ways similar to ours, was used to describe 
the probability of a cascade propagating from one network node to another in \cite{LarremorePRE12}.
Ref.~\cite{LarremorePRE12} is primarily motivated by avalanches of neurons firing, but also mentions blackouts, and analyzes the the statistics of the number of generations in a cascade as well as the number of failures.
One difference is that \cite{LarremorePRE12} restores nodes in the generation after they fail, which allows cascades to propagate through previously failed nodes and prolongs the cascades.
The ideas in~\cite{Hines:2013hicss} were extended in~\cite{Qi:2015} to build an interaction model of cascading that can reproduce 
statistical properties of simulated cascades.

This paper presents a new influence graph method, also based on the concepts in~\cite{Hines:2013hicss}, 
in which we take large amounts of data from cascade simulations and synthesize the data into a Markovian network model.
Importantly, the resulting model has a network structure through which cascades propagate locally, 
but that network structure is dramatically different from that of the original power network.
That is, the outages propagate along the influence graph, which is completely different than the graph topology of the grid.
We emphasize that any cascading failure simulation based on engineering principles can be used to produce the data needed to 
synthesize 
the influence graph.

\section{Method: Statistical modeling of non-local cascades}\label{method}

A cascading failure begins with one or more initiating events, typically component outages.
For example a tree falls into a transmission line, or an operator error results in one or more component outages.
Each initiating event will perturb the state of the network, which may result in excessive stress on other components.
Because of the laws of power flow, these stressed components may be topologically distant from the initiating events.
Excessive stress may cause one or more dependent outages, 
which may subsequently cause additional stress and additional outages.
Together this sequence is a cascading failure.

Prior work~\cite{DobsonPStrips12} has shown that one can gain substantial insight into the statistical properties of cascading failure by grouping sequences of component outages into generations, and looking at the growth (or propagation) rates among generations.
Typically, the first generation represents the exogenously caused initiating events.
Subsequent generations can be thought of as ``children'' of prior ``parent'' outage generations. 
In this branching process model, each generation of failures produces some number of dependent failures and the rate at which prior generation (parent) outages cause new (child) outages is known as the propagation rate, $\lambda$.
If $|Z_0|$ is the total number of outages (over many cascades) in the set of all initiating events over many cascades
and $|Z_1|$ is the number of outages in the first dependent generation, then the propagation rate from generation zero into the first generation is: $\overline{\lambda_0} = |Z_1|/|Z_0|$.
More generally:
\begin{align}
	\overline{\lambda_m} = \frac{|Z_{m+1}|}{|Z_{m}|} \label{lambda_m}
\end{align}

However, the approach of~\cite{DobsonPStrips12} simply counts outages without discriminating which component outages will result from a particular prior outage,
or how the components relate to other components. It is clear that component outages vary in their frequency and impact on other components. 
For example, the outage of a large transmission line carrying a large amount of power is likely to result in more dependent outages than the failure of a smaller line. 
Thus, it is useful to consider different components as having different propagation rates.
With this in mind, 
our model assumes that each component~$i$ produces a random number, $K_{i,m}$, of child outages according to the following conditional probability function:
\begin{align}
  f[k|i,m] =& \mbox{
  	$\Pr$[\,$k$ outages in generation $m+1$, given}\notag\\[-1mm]
	&\qquad\mbox{a single outage of $i$ in  generation $m$]}\label{flabel}
\end{align}
In this paper we assume that $f[k|i,m]$ is a Poisson distribution, with $\lambda_{i,m}$ representing the mean number of outages propagated by the outage of $i$ in generation(s) $m$. 

Finally, if component~$i$ outages and causes a dependent outage in the next generation, some components are more likely to outage in the next generation than others.
For the case of line outages, this increased likelihood comes from a number of factors including the way in which currents are redistributed (which can be estimated from line outage distribution factors) and the proximity of particular components to their tripping threshold.
Therefore, it is important to model the conditional probability of component $j$ failing, given the failure of $i$.
Let:
\begin{align}
	g[j|i,m]=&\mbox{ $\Pr$[\,$j$ fails in generation $m+1$ given }\notag\\[-1mm]
	         &\qquad\mbox{a single outage of $i$ in  generation $m$}\notag\\[-1mm]
			 &\qquad\mbox{and one outage in generation $m+1$]\label{glabel}}
\end{align}
For each generation $m$, $g[j|i,m]$ can be considered as the $i,j$ element of a matrix that defines a weighted, directed graph of influences among components.

Together $f$ and $g$ form a model, here referred to as an influence graph, that is Markovian in the sense that the outage probabilities depend only on the outages that occurred in the prior generation 
(see Sec.~\ref{simualting}).

\subsection{Estimating the parameters for $f$ and $g$ from data\label{estimating}}

In this paper we estimate the parameters for $f$ and $g$ from data obtained from many cascading failure simulations. 
While it may be possible in principle to design a method to estimate $f$ and $g$ directly from engineering information about the system, 
doing so would require extensive knowledge and difficult assumptions about how cascades propagate. By estimating the model's parameters from data our method can be applied to any  cascading simulation or model, so long as there is a discrete set of components that can fail, and these failures can be grouped into generations.

Let us assume then that a trusted engineering simulation model of cascading failure exists for a given network at a particular state, and that we can perturb this model with many random disturbances and thus produce a large amount of cascading failure sequence data.
The first step in building the influence graph is to group the outages from the sequence data into generations, typically by dividing the event sequences by finding pairs of events that are separated by some amount of time
(see~\cite{DobsonRA10,DobsonPStrips12}). 
After this grouping, we can 
let $Z_m^{(d)}$ represent the set of outages in generation $m$ of cascade $d$,
and $|Z_m^{(d)}|$ represent the number of outages within this set.

In order to estimate $f$ one first needs to decide the extent to which $f$ will be modeled to depend on the generation $m$.
For most of the results in this paper we provide separate estimates of $f$ for the initiating generation $f[k|i,0]$ and for the subsequent generations $f[k|i,1+]$. 
The rationale for this is that the $n-1$ security criterion results in networks being quite robust to initial outages, 
but after a cascade has already started outages propagate with a much higher probability.
For the simulated cascades used in this paper, this propagation rate does not change dramatically among the subsequent generations of dependent outages ($m\geq1$), denoted by 1+.
Thus, we describe the post-initiating-event propagation rates using the single (vector) distribution $f[k|i,1+]$. 

We assume that $f[k|i,m]$ follows a Poisson distribution, 
and need to estimate the Poisson parameter (the mean of $K_{i,m}$) for each component $i$ and each generation (or set of generations), $m$. 
Let $P_{i,m}$ represent the number of times (in a set of cascades) that each component~$i$ appears as a parent outage in generation(s) $m$, and $C_{i,m}$ denote the total number of ``effective children'' that result from the outage of component~$i$ in the respective generation(s).
More specifically, if $D_{i,m}$ is the set of cascades within which $i$ appears as a parent in generation(s) $m$, then:
\begin{align}
	P_{i,m} &= |D_{i,m}| \label{Pim} \\
	C_{i,m} &= \sum_{d \in D_{i,m}} \frac{|Z_{m+1}^{(d)}|}{|Z_{m}^{(d)}|} \label{Cim}
\end{align}
For example, if $i$ fails along with one other component $j$ in generation~0 of a particular cascade and generation~1 of that same cascade includes three outages, we would count $3/2=1.5$ additional ``effective children'' for each of $i$ and $j$, and thus add 1.5 to $C_{i,0}$ and $C_{j,0}$ and increment both $P_{i,0}$ and $P_{j,0}$.
After counting $P_{i,m}$ and $C_{i,m}$ for all of the cascades in a dataset, 
the Poisson parameter for $f[k|i,m]$ is $\lambda_{i,m} = C_{i,m}/P_{i,m}$.

When defined in this way, the weighted average (over $i$) of each $\lambda_{i,m}$ is equal to the overall propagation rate for the respective generation(s), $\overline{\lambda_m}$. 
For the specific case in which we estimate two sets of parameters for $\lambda_{i,0}$ and $\lambda_{i,1+}$
we get the following pair of relationships:
\begin{align}
	\overline{\lambda_0} &= \frac{ |Z_{1}| }{ |Z_{0}| } 
		= \frac{ \sum_{i=1}^n \lambda_{i,0} P_{i,0} } { \sum_{i=1}^n P_{i,0} } \\
	\overline{\lambda_{1+}} &= \frac{ \sum_{m=1}^{M-1} |Z_{m+1}| }
	                                { \sum_{m=1}^{M-1} |Z_{m}| } 
		= \frac{ \sum_{i=1}^n \lambda_{i,1+} P_{i,1+} } { \sum_{i=1}^n P_{i,1+} } 
\end{align}
where $M$ is the maximum generation index $m$ over all of the cascades in the dataset.

To build the graph of inter-component influences, $G$, 
we again iterate through each cascade $d$ and generation $m$.
For each $d$ and $m$ in which $i$ occurs as a parent and 
$j$ occurs as a child in the next generation ($m+1$), we add the fraction 
$1/|Z_{m}^{(d)}|$
to a counter $g_{c}[j|i]$.
Finally, the individual elements of $G$ are estimated as follows:
\begin{align}
	g[j|i] = \frac{ g_{c}[j|i] }{ \sum_{j=1}^n g_c[j|i] } 
\end{align}
Normalizing in this way ensures that $g[j|i]$ acts as a conditional probability, 
as per our definition in~(\ref{glabel}), such that
$\sum_{j=1}^n g[j|i] = 1$.
Note that in this paper we assume that $g$ does not change with generation~$m$.
The reason for this is that even with data from many simulated cascades, only a fairly limited number of observations are available for most of the potential sequence pairs $i \rightarrow j$.	

\subsection{6-bus illustration}\label{6bus-results}

\begin{figure}[H]
	\centering \includegraphics[width=0.8\columnwidth,height=2in]{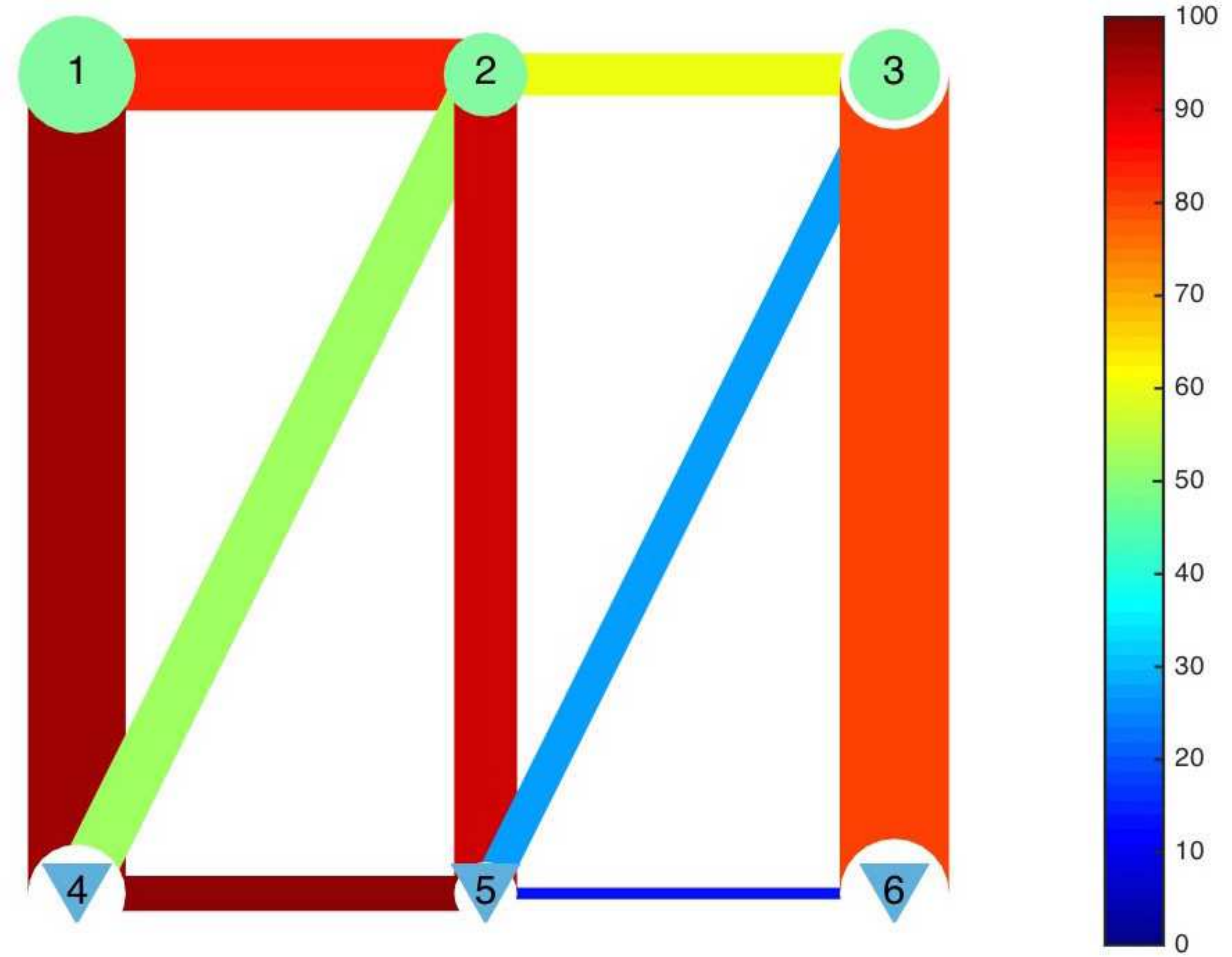}
	\vspace{-.1in}
	\caption{%
	The modified 6 bus test case with three generators (circles) and 3 loads (triangles).
	Line widths indicate absolute line loading (MW) and colors indicate precent of rated loading. 	
	}
	\label{6bus}
\end{figure}

\begin{figure}[H]
	\centering \includegraphics[width=0.8\columnwidth]{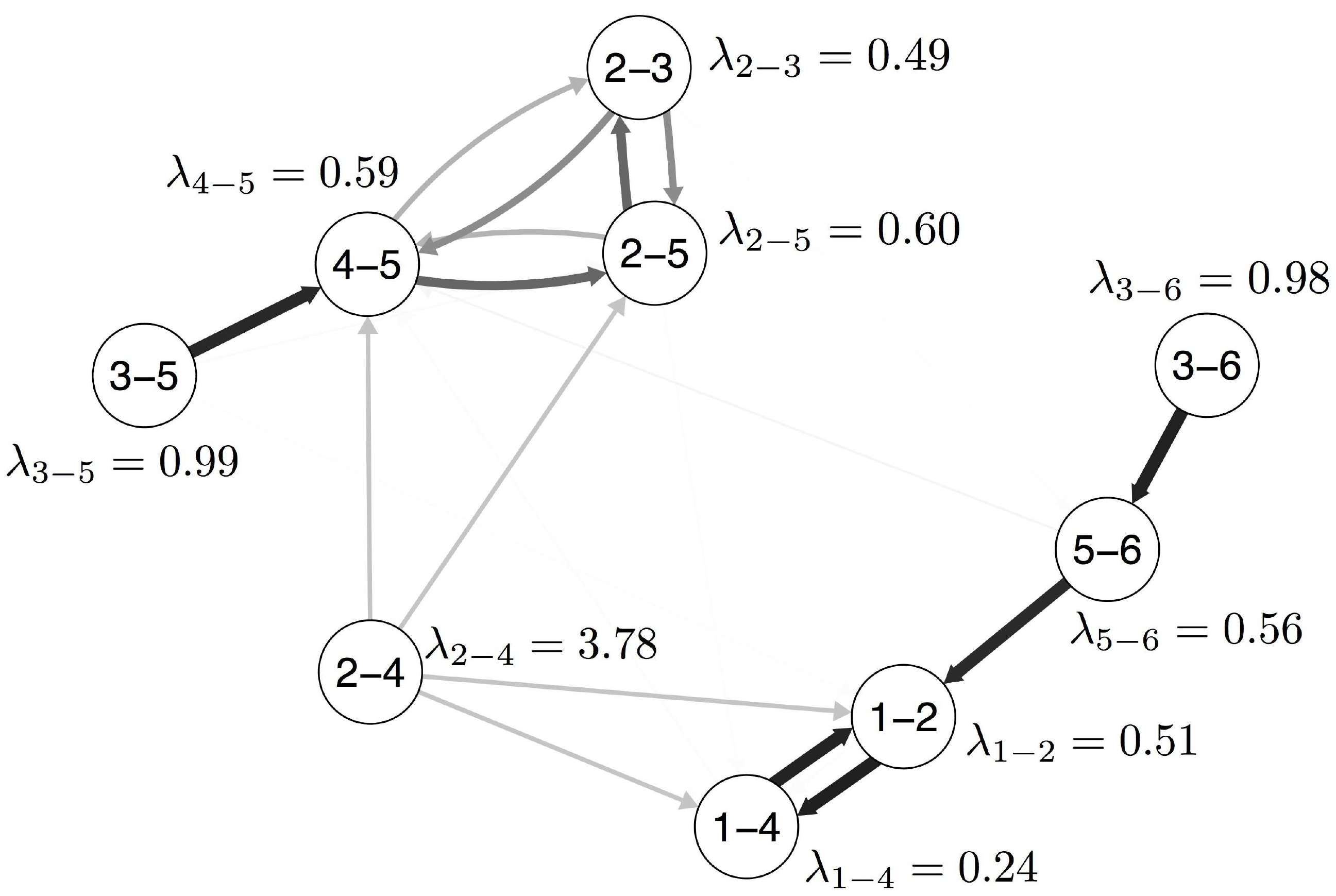}
	\vspace{-.1in}
	\caption{Illustration of the influence graph for the 6-bus test case. 
	The edge weights represent the conditional probability of one line outage propagating associated line outages.
	}
	\label{6bus-G}
\end{figure}

To illustrate the process of building $f$ and $g$, this section describes the formation of the influence graph for a slightly modified version of the Wood and Wollenberg 6-bus test case~\cite{Wood:1996} 
(see Fig.~\ref{6bus}).
After removing two transmission lines (for graphical clarity), 
all of the pre-contingency line flows were below the rated limits, but the system was not initially $n-1$ secure.
To generate cascading outage data, 1000 sets of one or more initiating outages were randomly generated assuming that each of the nine transmission lines 
had an equal outage probability of $p_0=1/100$.
Dependent outage sequences were generated using the DCSIMSEP cascading failure model from~\cite{Eppstein:2012,Rezaei:2015rc}.
These sequences were then grouped into generations by separating outages that were distant in time by at least $\Delta t = 5$ seconds.
Note that this is a relatively small value for $\Delta t$ compared to previous branching process applications (e.g.,~\cite{DobsonPStrips12}). Since the 6-bus case is very small, and used here only to illustrate the influence graph concept, choosing a small $\Delta t$ seemed appropriate.
Given data separated into generations we computed a single distribution for $f$, as shown in Table~\ref{6busStats}.
Similarly the parameters for $g$ were computed as described in Sec.~\ref{estimating}.
The results for $f$ and $g$ are illustrated in Fig.~\ref{6bus-G}.

A number of observations result from inspection of $f$ and $g$. 
From $f$ we see that one of the nine lines (line 2-4) has a much higher propagation rate than the other eight, a clear sign of its critical importance to this particular system.
We also find two different common paths of cascading. 
One path includes the set of lines \{3-6,5-6,1-2,1-4\} and 
a second includes \{3-5,4-5,2-5,2-3\}. 
The outage of 2-4 can propagate a subset of either of these two paths.

\begin{table}[ht]
	\setlength\tabcolsep{0.05in}
	\caption{Parameters for $f$ for the 6-bus test case
	\label{6busStats}}
	\centering
	\begin{tabular}{cccccccccc}
	\hline 
	Line: & 1-2 & 1-4 & 2-3 & 2-4 & 2-5 & 3-5 & 3-6 & 4-5 & 5-6 \\
	\hline 
	$P_{i}$ & 412 & 416 & 401 & 106 & 498 & 100 & 104 & 503 & 193 \\
	$C_{i}$ & 208.5 & 98.5 & 195.5 & 401 & 298.5 & 99 & 102 & 297.5 & 107.5 \\
	$\lambda_{i}$ & 0.51 & 0.24 & 0.49 & 3.78 & 0.60 & 0.99 & 0.98 & 0.59 & 0.56 \\
	\hline 
	\end{tabular}
\end{table}

\section{Simulating cascading failures given $f$ and $g$}\label{simualting}

Once formed, $f$ and $g$ can be used to rapidly generate many synthetic cascades, which have statistical properties similar to those of the original data.

\begin{figure}[t]
	\includegraphics[width=1.0\columnwidth]{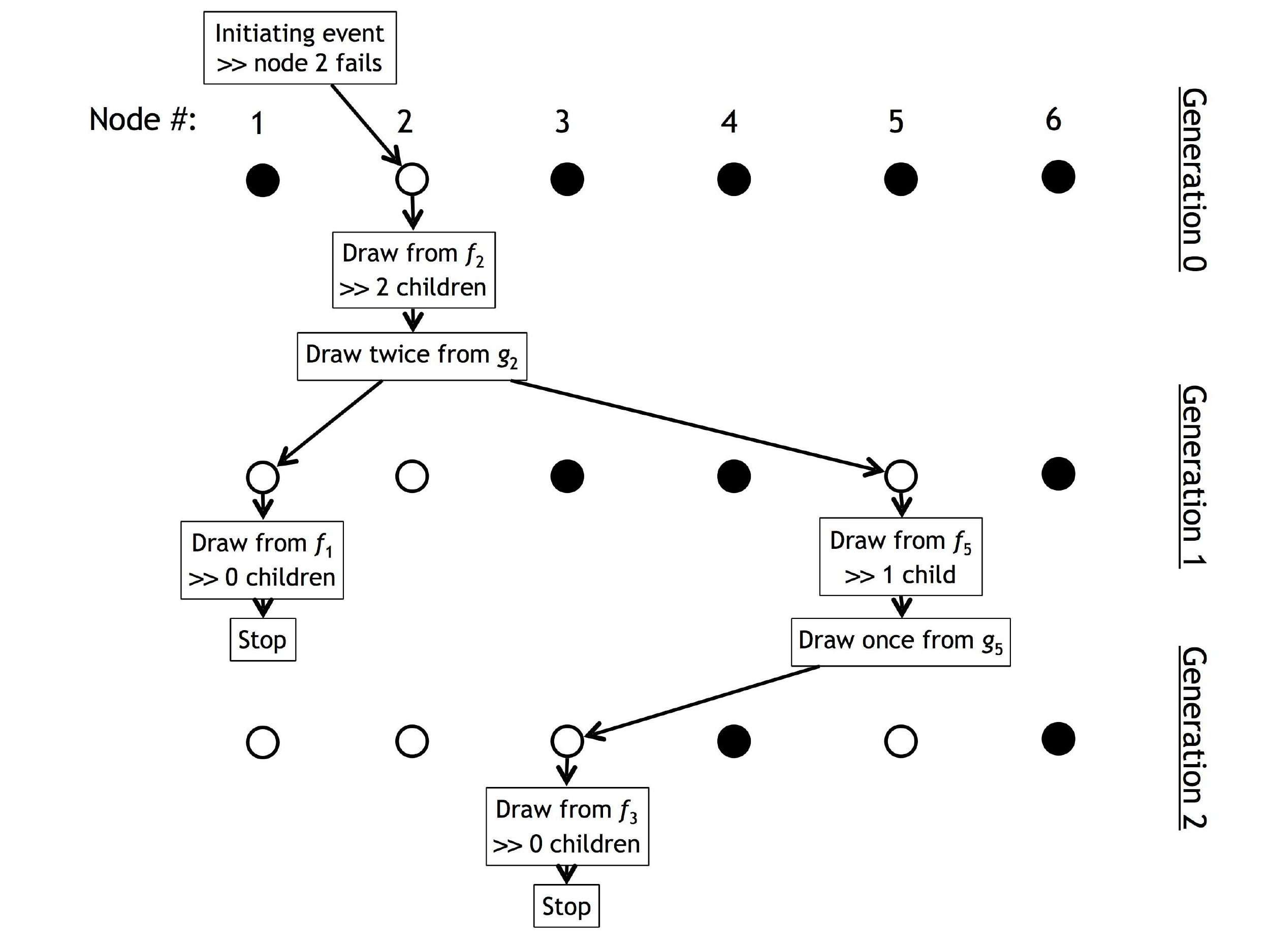}
	\vspace{-.3in}
	\caption{Illustration of an influence graph simulation for a small network with six components. 
	Initially node~2 fails due to some exogenous cause.
	Based on a draw from $f_2$, this failure causes $k=2$ dependent failures in the next generation.
	Drawing twice from $g[j|i=2]$ results in nodes 1 and 5 failing in the next generation.
	In generation 1, the failure of node 1 does not produce any additional children, 
	but the failure of 3 produces one ``child'' failure (node~3).
	Finally, the failure of node~3 does not produce any children, thus ending the cascade.
	}
	\label{igraph_cartoon}
\end{figure}

To simulate the influence graph (see Fig.~\ref{igraph_cartoon}), we start with a large set of initiating events.
These can be produced by a variety of sampling methods, such as Monte-Carlo or complete enumeration of a plausible contingency list (e.g., all $n-2$'s).
For Monte-Carlo sampling, if initiating outages are assumed to be independent, 
then this can be a simple vector of failure probabilities for each of $n$ components. 
Given the chosen sampling method, the influence graph can be simulated as follows:
\begin{enumerate}
	\item Initialize the cascade index: $d=1$
	\item Initialize the generation index: $m=0$ \label{init_gen}
	\item Produce a set of exogenous initiating outages $Z_{0}^{(d)}$ via the chosen sampling method.
	\item For each outage $i \in Z_{m}^{(d)}$, do the following: 
	 (a) Determine how many child outages $\kappa$ result from $i$ by sampling from $f[k|i,m]$.
	 (b) Determine which outages result from outage $i$ by sampling from $g[j|i]$~$\kappa$ times. 
	 This sampling is done using Bernoulli trials, 
	 such that for each trial $\{1 \ldots \kappa\}$ component $j$ will fail if $g[j|i]>r$, 
	 where $r$ is a uniformly distributed random number $0 \leq r \leq 1$.
	 The result is a set of outages for the next dependent generation, $Z_{m+1}^{(d)}$.
	 \label{choose_children}
	\item If $Z_{m+1}^{(d)}$ includes at least one outage, then increment the generation index $m=m+1$ and continue from Step~\ref{choose_children}.
	\item Otherwise, increment the cascade counter $d=d+1$ and continue from Step~\ref{init_gen}.
\end{enumerate}
Note that a given component $j$ may be selected for failure more than once. 
Since a component cannot fail multiple times (and restoration is not included in this model), 
this means that the total number of new child outages may be slightly less than~$\kappa$. 

\subsection{Simulated cascades in a larger test case}\label{polish-results}

\begin{figure}[ht]
	\includegraphics[width=1.0\columnwidth]{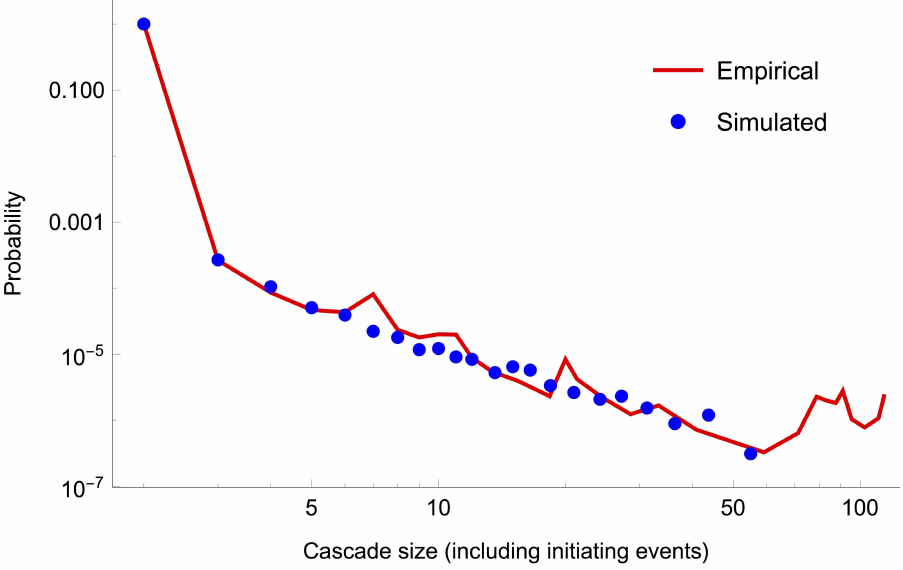}
	\caption{Probability distribution  of cascade sizes,
	measured by the number of branches failed, 
	for the original DCSIMSEP simulation data (empirical), 
	and for the influence graph (simulated) data. 
	Data are binned so that each bin contains at least 26 observations.
	\label{polish-sizes}}
\end{figure}

To illustrate the influence graph method, we produced a dataset of cascading outages by simulating the impact of each $n-2$ transmission branch (transformer or line) outage in an $n-1$ secure version of the winter peak Polish case available with MATPOWER~\cite{MATPOWER:2011}, using the same simulator (DCSIMSEP) and case data as in~\cite{Eppstein:2012,Rezaei:2015rc}.
The test case has $n=2896$ branches, which resulted in $n(n-1)/2=4\,191\,960$ initiating $n-2$ contingencies, of which $3170$ resulted in at least one dependent outage. 
Fig.~\ref{polish-sizes} shows the distribution of cascade sizes for this dataset.
\begin{figure}[ht]
	\includegraphics[width=1.0\columnwidth]{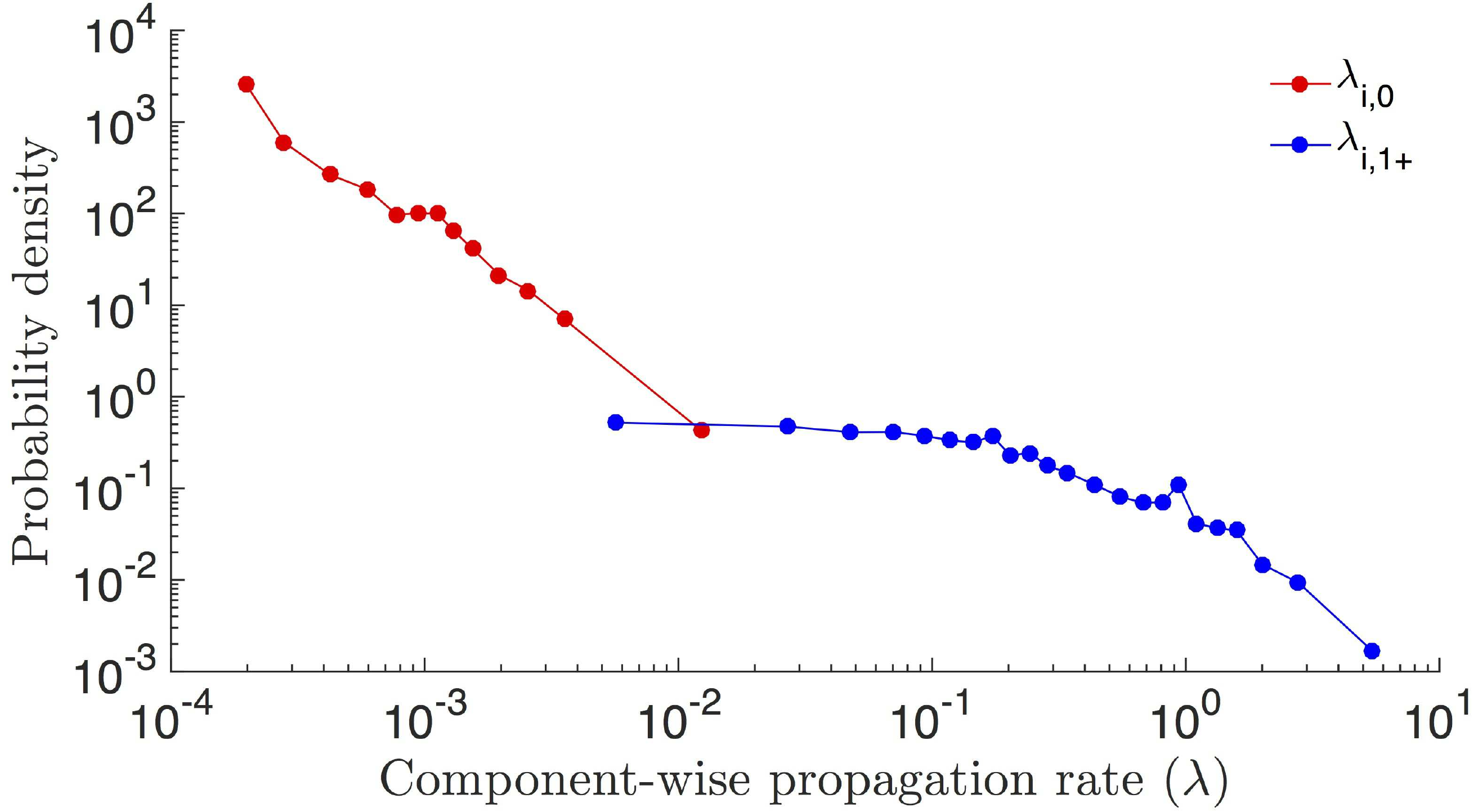}
	\vspace{-.2in}
	\caption{Empirical probability density functions for the component-wise propagation rates in the initiating generation, $\lambda_{i,0}$, 
	and the subsequent generations, $\lambda_{i,1+}$.
	Data are binned as in Fig.~\ref{polish-sizes}.}
	\label{f_pdf}
\end{figure}

The outage data were subsequently separated into generations by assigning all of the initiating outages to the first generation and distributing the dependent events into subsequent generations by assigning 
events that were at least 30 seconds apart~\cite{DobsonPStrips12} 
to different generations. 
We chose this value for $\Delta t$ based on the fact that fast transients and auto-recloser actions are typically completed within 30 seconds.
We also tested $\Delta t=15$s and $\Delta t=60$s, and found that the parameters for $f$ were not substantially altered within this range.
After separating the outage data into generations, the parameters for $f$ and $g$ were constructed as described in Sec.~\ref{estimating}.
Figure~\ref{f_pdf} shows the distribution of values for $\lambda_{i,0}$ and $\lambda_{i,1+}$.
Notably, both distributions are heavy tailed; for the majority of components $\lambda_{i,m}=0$, but a few components tend to produce a much larger number of ``child'' outages. 

Finally, we simulated artificial cascades by applying each of the $4\,191\,960$ possible $n-2$ contingencies with the resulting influence graph.
Figure~\ref{polish-sizes} shows the resulting probability distribution of cascade sizes, measured as the total number of outages in each cascade (including the initiating outages), for the empirical data and the simulated data from the influence graph. 
The influence graph method matches the empirical data quite well, with the exception that the influence graph does not reproduce the frequency of the very longest cascades observed in the empirical data. 
Otherwise the match between the simulated and empirical data is quite good.

\begin{figure}[H]
	\centering
	\includegraphics[width=0.9\columnwidth]{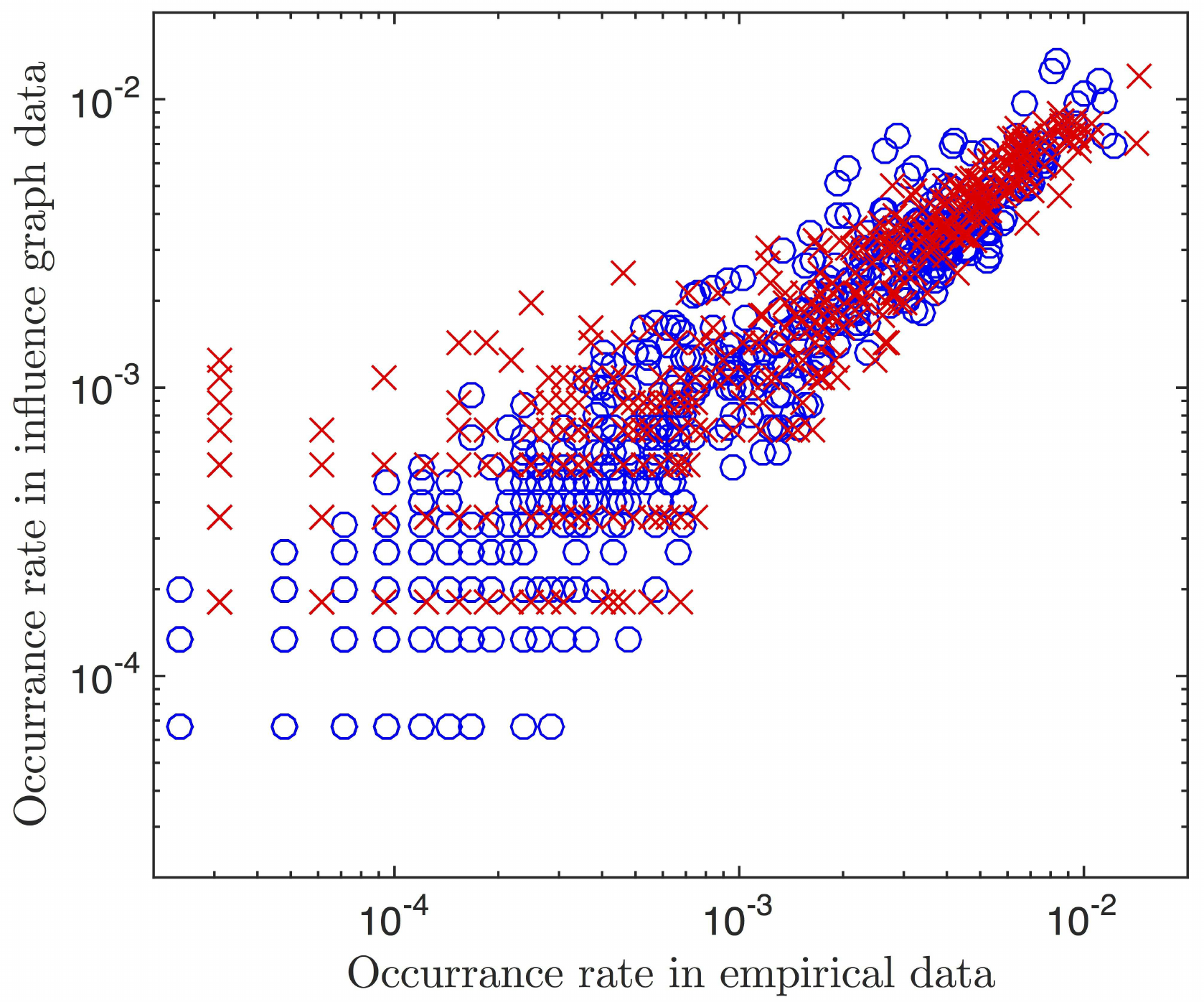}
	\caption{Comparison of the outage rate of components in the influence-graph simulated data and in the original cascading failure data. 
	Circles ($\color{blue} \circ$) show the rates for all dependent events and $\color{red} \times$ shows only dependent events that occurred in generation $m=4$ and later. 
	}
	\label{scatter}
\end{figure}

Another way to compare the influence graph data to the original data is to look at the frequency with which particular components appear in the two datasets.
One would expect a valid model of the original data to show that component $i$ outages with a frequency that is similar to its outage rate in the original data.
To evaluate if this is indeed the case, Figure~\ref{scatter} compares the outage rates (the number of dependent outages of $i$ divided by the total number of dependent outages) in the simulated data and in the empirical data. 
The match between these two rates provides further evidence for the validity of this approach.

\section{Extracting useful information from the influence graph}
\label{useful}

Once the influence graph ($f$ and $g$) is built from data the results can shed valuable insight into the general properties of cascading in a particular system.

In order to do so, we first combine $f$ and $g$ into a single graph $H$ that captures the relative strength of the influences among the component outages.
Consider that 
in an arbitrary generation~$m$ of influence-graph-simulated cascade~$d$, 
component~$i$ fails alone ($Z_m^{(d)}=\{i\}$) 
and a random draw from $f[k|i,m]$ indicates that there should be (about) $K=k$ failures in the next generation. 
Then the conditional probability that a particular component $j$ fails in the next generation ($m + 1$), given that $i$ failed in generation $m$ and that generation $m + 1$ includes exactly $k$ failures is:
\begin{align}
	\Pr( j | \{i,k\} ) =  1 - (1 - g[j|i] )^{k} \label{pr_j}
\end{align}
Let $h_{i,j,m}$ be the conditional probability that a particular component $j$ fails in generation $m+1$ given that $i$ failed in $m$, over all values of $k$. 
$h_{i,j,m}$ can be computed from~(\ref{pr_j}) using the assumption that $K$ is a Poisson random variable:
\begin{align}
	h_{i,j,m} &= \sum_{k=0}^{n-1} \Pr( j | \{i,k\} ) f[k|i,m] \\
		      &\sim \sum_{k=0}^{\infty}\left(1-(1-g[j|i])^{k}\right) f[k|i,m]  \\
			  &= \sum_{k=0}^{\infty}\left(1-(1-g[j|i])^{k}\right)
			     \frac{\lambda_{i,m}^{k}}{k!}e^{-\lambda_{i,m}}\\
              &=  1-e^{-\lambda_{i,m}}e^{\lambda_{i,m}(1-g[j|i])} \\
			  &=  1-e^{-\lambda_{i,m}g[j|i]} \label{hij}
\end{align}
Thus defined, $h_{i,j,m}$
can be thought of as the $i,j$ element of a matrix $\mathbf{H}_m$
that combines $f$ and $g$.
$\mathbf{H}_m$ has the properties of a weighted adjacency matrix for a directed graph.
Since, for our larger test case, we defined two different distributions for $f[k|i,m=0]$ and $f[k|i,m=1+]$, we end up with two different matrices $\mathbf{H}_0$ and $\mathbf{H}_{1+}$ that, respectively, describe the propagation of the initiating contingency and the subsequent dependent events.
Note that the nodes of the influence graph do not represent system states, and thus
$\mathbf{H}$ differs from a typical Markov chain transition matrix. 
In particular, $\mathbf{H}$ is not a stochastic matrix because the events comprising one of the components outaging after component $i$ outages are neither exclusive nor exhaustive; it is routine that no components or several components outage after component $i$ outages.


\subsection{Using $\mathbf{H}$ to find critical components}

A particularly important question in the study of cascading failure risk is that of identifying critical components,
the failure of which could result in particularly large cascading failures.
Or, more practically, finding components that could be improved in some way to substantially reduce blackout risk.
Prior work~\cite{Rezaei:2015rc,kaplunovich2014statistical} has suggested that some components, when they fail as a part of a multiple initiating contingency, contribute orders of magnitude more to blackout risk, relative to the average component.
Here we suggest a method for using the information contained in $\mathbf{H}_0$ and $\mathbf{H}_{1+}$ to find those components that could propagate large cascading failures if they fail \emph{during} a cascade, as opposed to during the initiating contingency.
This type of information could be useful 
to power system planners in identifying components that should be prioritized for more thorough vegetation management or upgraded protection systems, or 
to operators in making adjustments to the dispatch to reduce blackout risk by decreasing line loadings.

We start by defining a vector of independent binary (Bernoulli) random variables, $\mathbf{s}_0$, that describes the space of possible initiating contingencies.
$\mathbf{s}_0$ is defined such that $s_{i,0} = 1$ indicates that component~$i$ outages as a part of the initiating contingency and $s_{i,0} = 0$ means that $i$ did not initially fail.
$p_{i,0}$ is the probability that $s_{i,0} = 1$ and 
thus $p_{i,0}$ also is also the expected value of $s_{i,0}$.
More generally, let $p_{i,m}$ represent the probability that $i$ outages in generation $m$, and $\mathbf{p}_m$ be the column vector of these probabilities.

Let us now use $\mathbf{H}_m$ and the outage probabilities $\mathbf{p}_m$ to find the probability that a particular component $j$ outages in generation $m+1$.
$h_{i,j,m}$ gives us the probability that $j$ outages in generation $m+1$, given that $i$ outaged in generation $m$.
The probability $p_{j,m+1}$ that $j$ outages in generation $m+1$ is the probability of the union of all the ways that $j$ might have failed.
If the individual interaction probabilities in $\mathbf{H}$ are small and the ways are independent (or disjoint), then we can neglect the higher order probabilities (such as the probability that $j$ failed due to the combination of $i$ and $k$), and obtain the following approximation for the probability that $j$ fails in generation $m+1$:
\begin{align}
	p_{j,m+1} &\cong \sum_{i=1}^n \Pr[ j(m+1)|i(m) ] \Pr[ i(m) ] \\
	 &= \sum_{i=1}^n h_{i,j,m} p_{i,m} \label{eq:pm}
\end{align}
As a result, the outage probabilities in the first and subsequent generations can be estimated by simple matrix multiplication:
\begin{align}
	\mathbf{p}_1^\intercal     &= \mathbf{p}_0^\intercal \mathbf{H}_0  \label{eq:p1} \\
	\mathbf{p}_{m+1}^\intercal &= \mathbf{p}_m^\intercal \mathbf{H}_{1+}\quad,\  m \geq 1. \label{eq:pm+}
\end{align}
After a long cascade, the probability of each component having failed at some point during a cascade (the vector $\mathbf{a}$) can be found as follows:
\begin{align}
	\mathbf{a}^\intercal &\triangleq \sum_{m=0}^\infty \mathbf{p}_m^\intercal 
	= \mathbf{p}_0^\intercal + \mathbf{p}_0^\intercal \mathbf{H}_0 \sum_{m=0}^\infty \mathbf{H}_{1+}^m \label{eq:a_before}
\end{align}
Eq.~(\ref{eq:a_before}) will evaluate to a finite quantity, so long as the absolute values of the eigenvalues of $\mathbf{H}_{1+}$ are less than 1, which holds for  our test cases.
Since  $\sum_{m=0}^\infty \mathbf{H}_{1+}^m = (\mathbf{I} - \mathbf{H}_{1+})^{-1}$, Eq.~(\ref{eq:a_before}) can be rewritten more simply as:
\begin{align}
	\mathbf{a}^\intercal &= \mathbf{p}_0^\intercal + \mathbf{p}_0^\intercal \mathbf{H}_0 (\mathbf{I} - \mathbf{H}_{1+})^{-1}  \label{eq:a}
\end{align}
This is a useful expression as it enables us to quickly understand which components are most likely to be involved in a cascading failure.
Also, the sum of $\mathbf{a}$ gives an estimate of the expected size of the cascades that could result from the vector of initiating probabilities $\mathbf{p}_0$.

Now we turn our attention to using  (\ref{eq:a}) to estimate the impact of design changes on cascading failure propagation.
Let us assume that we know that if we 
upgrade component~$j$ (perhaps by replacing line $j$'s distance relays with current differential relays, or improving vegetation management on its transmission corridor) we can reduce the probability of line $j$ outaging due to an overload, in response to other outages in the system.
We can represent this change by reducing the probabilities in the $j^\mathrm{th}$ column of $\mathbf{H}_0$ and $\mathbf{H}_{1+}$ by subtracting column perturbation vectors 
$\delta_0$ and $\delta_1$ from $\mathbf{H}_0$ and $\mathbf{H}_{1+}$, respectively.
Then the result of the perturbations to $\mathbf{H}_0$ and $\mathbf{H}_{1+}$ can be estimated by calculating 
\begin{align}
	{\mathbf{a}'_j}^{\intercal} &= 
	\mathbf{p}_0^\intercal + 
	\mathbf{p}_0^\intercal (\mathbf{H}_0 - \delta_0 \mathbf{e}_j^\intercal ) 
	(\mathbf{I} - \mathbf{H}_{1+} + \delta_1 \mathbf{e}_j^\intercal )^{-1} \label{a_prime}
\end{align}
where 
$\mathbf{e}_j$ is an indicator vector with 1 in the $j^\mathrm{th}$ element and 0 elsewhere.

However, what we really care about, if we want to estimate the impact of potential upgrades, is the \emph{change} in $\mathbf{a}$ that results from the perturbations $\delta_0$ and $\delta_1$, not the absolute value of $\mathbf{a}_j'$. 
This change is computed by subtracting (\ref{a_prime}) from (\ref{eq:a}):
\begin{align}
	&\Delta {\mathbf{a}_j}^\intercal = \mathbf{a}^\intercal - {\mathbf{a}'_j}^\intercal = \\
	&\, \mathbf{p}_0^\intercal 
	 	\left( 
		 \mathbf{H}_0
		 ( \mathbf{I} - \mathbf{H}_{1+} )^{-1}
	   - ( \mathbf{H}_0 - \delta_0 \mathbf{e}_j^\intercal ) 
	     ( \mathbf{I} - \mathbf{H}_{1+} + \delta_1 \mathbf{e}_j^\intercal )^{-1} 
		 \right)
		 \notag
\end{align}
Applying the Sherman-Morrison formula~\cite{sherman1950adjustment} for rank-1 updates of an inverted matrix and simplifying leads to:
\begin{align}
	\Delta {\mathbf{a}_j}^\intercal =& \,	 
		\mathbf{p}_0^\intercal 
		\delta_0 \mathbf{e}_j^\intercal ( \mathbf{I} - \mathbf{H}_{1+} )^{-1} + 
		\label{eq:d_ai} \\
		& \mathbf{p}_0^\intercal 
		(\mathbf{H}_0 - \delta_0 \mathbf{e}_j^\intercal)
		\frac
			{ (\mathbf{I} - \mathbf{H}_{1+})^{-1} \delta_1 \mathbf{e}_i^\intercal (\mathbf{I} - \mathbf{H}_{1+})^{-1} }
			{ 1 + \mathbf{e}_j^\intercal (\mathbf{I} - \mathbf{H}_{1+})^{-1} \delta_1 } 
		\nonumber 
\end{align}
Finally we can sum $\Delta \mathbf{a}_j$ to define  the overall impact $\alpha_j$ of modifications $\delta_0$ and $\delta_1$ 
on line $j$:
\begin{align}
	\alpha_j = \sum_{k=1}^{n} \Delta a_{j,k} \label{alpha_j}
\end{align}
By computing $\alpha_j$ for each  line $j$ from the parameters of the influence graph
we can quickly estimate the impact of potential modifications without the need for extensive simulations of these modifications.
This metric (\ref{alpha_j}) allows us to quickly compute, from data, the relative importance (or ``criticality'') of particular components in a power system.


\subsection{Critical components in the 6-bus case}

\begin{figure}[H]
	\vspace{-.1in}
	\centering \includegraphics[width=0.75\columnwidth]{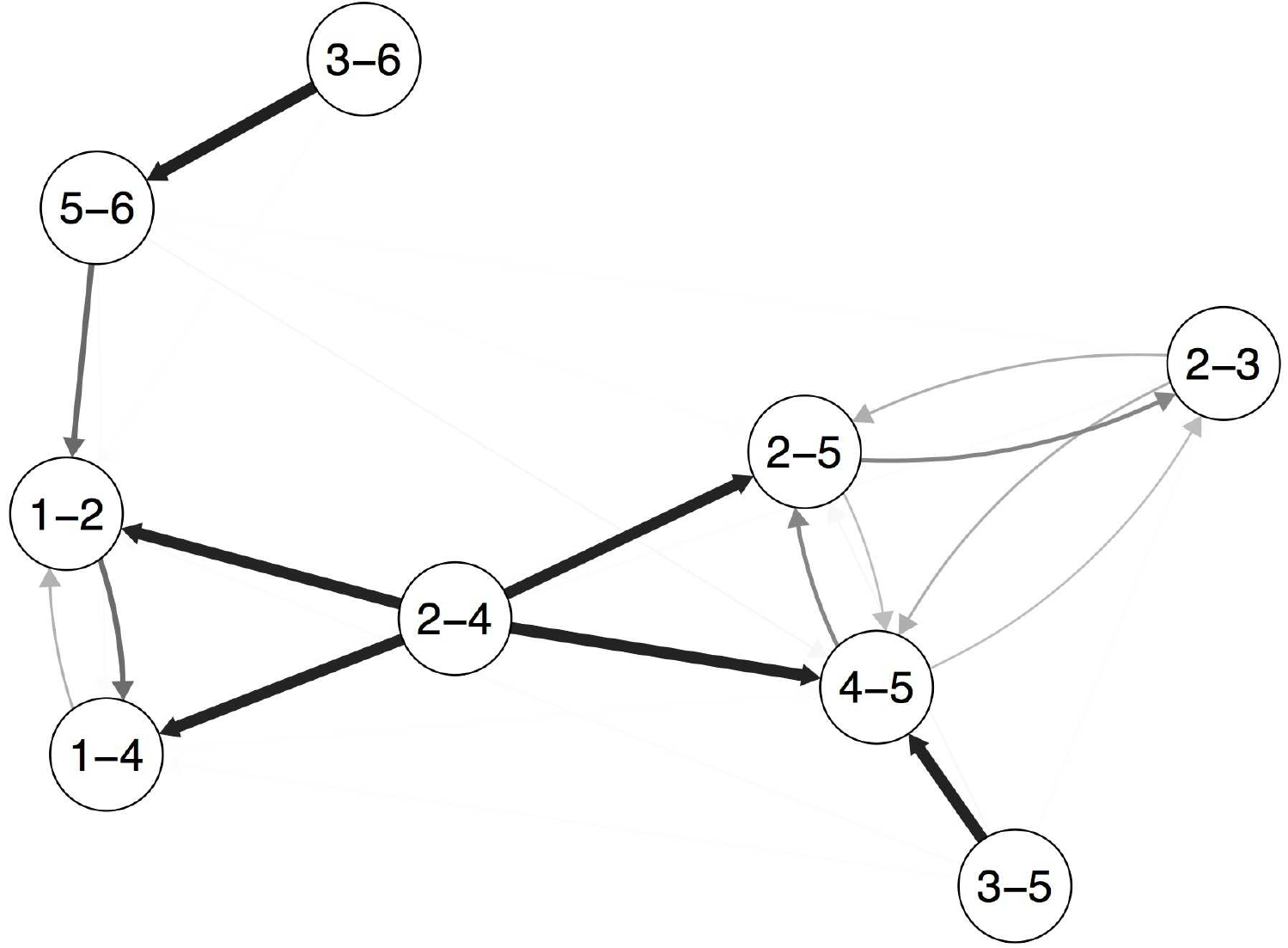}
	\vspace{-.1in}
	\caption{Illustration of the combined influence matrix $\mathbf{H}$ for the 6 bus case.
	Edge weights show the probability of an outage at the origin node resulting in 
	an outage at the destination node.
	}
	\label{6bus_glam}
\end{figure}

To illustrate the result of combining $f$ and $g$ to produce a single influence graph, as defined in (\ref{hij}), 
Fig.~\ref{6bus_glam} shows $\mathbf{H}$ for the 6-bus test case presented in Sec.~\ref{6bus-results}.
(For this small case we assume that $\mathbf{H} = \mathbf{H}_0 = \mathbf{H}_{1+}$.)
This figure clearly shows the importance of line (2-4), since it has a high likelihood of initiating four subsequent outages, 
which can, in turn, propagate additional outages.
However, (2-4) only appears as a initiating event, as indicated by the fact that this node has zero in-degree; other outages do not result in the outage of (2-4).
As a result, modifying the probability of (2-4) failing endogenously (within a cascade) will not have an effect on cascade sizes.
On the other hand, the failure of either (2-5) or (4-5) can propagate as many as three additional outages, making these components candidates for potential upgrades.

\begin{figure*}[t]
	\centering \includegraphics[width=1\textwidth]{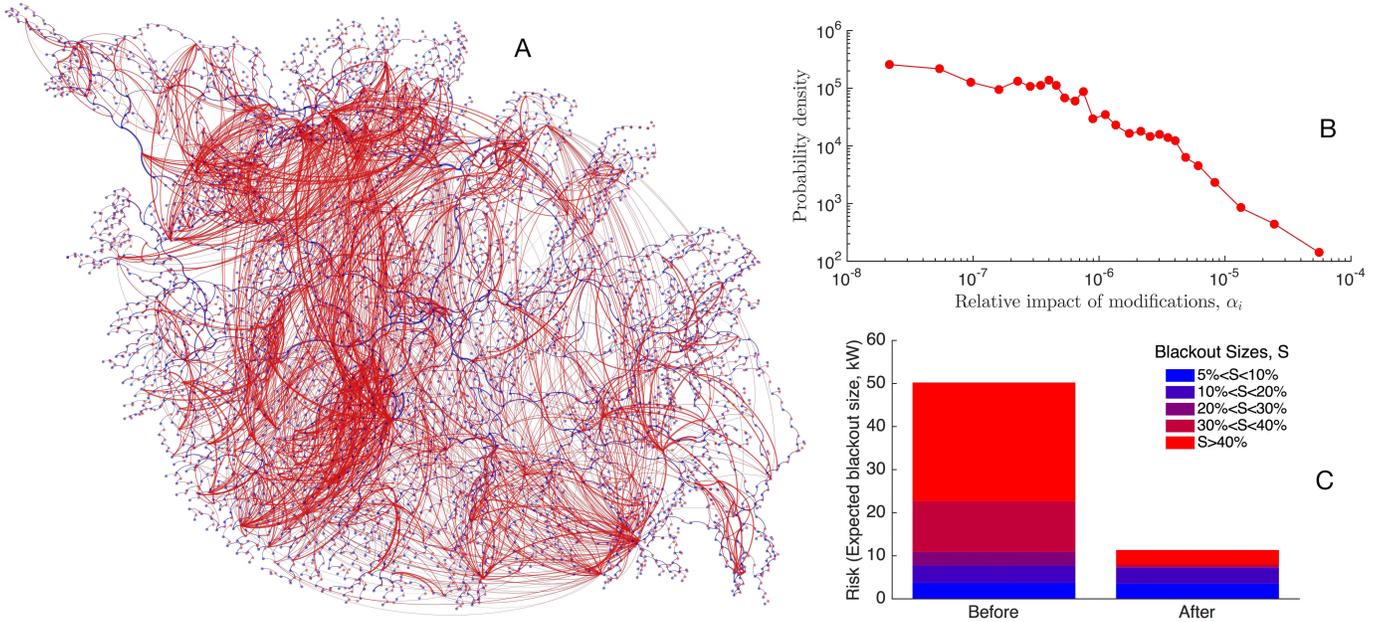}
	\caption{%
	(A) Illustration of the combined $\mathbf{H}_{1+}$ for the Polish test case.
	Blue nodes/links show the physical power grid, and red nodes/links show the influences among branch outages. 
	Link thicknesses indicate power flow magnitudes (blue) and influence probabilities (red).
	Note that for graphical clarity very small influences are not depicted in this figure.
	(B) The probability distribution of the metric $\alpha_i$ for the same test case. 
	(C) The risk from cascading blackouts initiated by $n-2$ contingencies in the original system before and after increasing flow limits for the ten most critical lines.
	}
	\label{polish_igraph}
\end{figure*}

These qualitative observations can be made quantitative by applying equations (\ref{eq:a}) and (\ref{alpha_j}) to $\mathbf{H}$ for the 6-bus case. 
The importance of component (2-4) to the system can be observed by computing (\ref{eq:a}) after setting $\mathbf{p}_0$ to indicate that only (2-4) outages initially, with probability 1.
In this case, the expected cascade size from (\ref{eq:a}) is 
$\mathbf{a}^\intercal \mathbf{1} = 5.16$.
An initial outage of the other lines in the network is expected to produce much smaller cascades. For all of the remaining lines $\mathbf{a}^\intercal \mathbf{1} < 2.28$.
Similarly, we can estimate the effect of potential upgrades to this system by computing  $\alpha_j$ in (\ref{alpha_j}).
To do so we choose $\delta_0 = \delta_1 = \delta$ equal to 1/2 of the $j^\mathrm{th}$ column of $\mathbf{H}$ for each of the 9 lines in this network, and set $\mathbf{p}_0$ to a vector of initiating probabilities with $p_{0,i} = 0.001, \, \forall i$.
The results (unsurprisingly) tell us that modifications to (2-5) and (4-5) will be most beneficial, with 
$\alpha_{\mathrm{(2-5)}} = 0.021$ and 
$\alpha_{\mathrm{(4-5)}} = 0.022$.
$\alpha_j < 0.015$ for the remaining components.
Since they have no in-degree (no inward links) $\alpha_{i}=0$ for (2-4), (3-5) and (3-6).

\subsection{Polish test case results}

To test these ideas for a larger network, we computed $\mathbf{H}_0$ and $\mathbf{H}_{1+}$ 
from $f[k|i,0]$, $f[k|i,1+]$ and $g[j|i]$, 
and studied the properties of the resulting influence graph.
Figure~\ref{polish_igraph}A shows $\mathbf{H}_{1+}$ for the Polish test case.
While the detailed structure is difficult to visualize, what is clear is that the influence graph has a topological structure that is distinctly different from that of the underlying physical infrastructure of buses and transmission lines.
We argue that this difference in structure at least partially explains the substantial differences between the vulnerability implications that one obtains from power grid simulations and from simple topological metrics computed from the physical network, as observed in~\cite{Hines:2010}.

In order to test our metric for component criticality, we computed $\alpha_j$ for each of the $n = 2896$ branches.
To do so all of the initiating event probabilities were set to $p_{i,0} = 1/8760$, based on the assumption that all components outage at a rate of about 1 outage per year.  Note that this assumed outage rate has no effect on our conclusions about which component is the most critical, since the metric in (\ref{eq:d_ai}) is scaled uniformly by $\mathbf{p}_0$.
Secondly, we computed the perturbations $\delta_0$ and $\delta_1$ based on an assumed 50\% reduction in the propagation rate. 

Figure~\ref{polish_igraph}B shows the empirical probability distribution for the resulting criticality metric.
This distribution clearly shows a heavy-tailed (nearly power-law) pattern:
while the vast majority of the potential upgrades have little-to-no effect 
($\alpha_j<10^{-7}$ for 83\% of the components), a few components have nearly three orders of magnitude greater impact than this.

To evaluate the extent to which this information could be useful in a planning context, we performed the following calculation.
We took the ten lines that appeared to be most critical from $\alpha_j$ and then doubled their flow limits used in the cascading failure simulation, without changing the pre-contingency dispatch or power flows.
Something similar to this type of increased flow limit could be accomplished by disabling backup (e.g., Zone 3) relaying systems for these critical lines, by replacing simple distance relays with more sophisticated current-differential relaying systems, or by re-conductoring.

After increasing the line flow limits, we re-simulated the entire set of $n-2$ contingencies and recomputed the cascading failure sizes.
While this change does not substantially change the frequency of small cascades, the impact on the frequency of very large cascades, which have the greatest impact on blackout risk, is dramatic.
The probability of a cascade that includes 50 or more outages goes down by 94\% in the modified system.
As shown in Fig.~\ref{polish_igraph}C, this relatively small modification reduces the risk of large cascading blackouts (those resulting in 5\% or more load shedding) by about 80\%. 

\looseness=-1
Another way to compare these two cases is to look at the propagation rate $\overline{\lambda_{1+}}$ in the system before and after modification.
In the original data the propagation rate was 
$\overline{\lambda_{1+}} = 0.92$.
In the modified system, the propagation rate is reduced to
$\overline{\lambda_{1+}} = 0.79$.
This relatively small difference can have a large impact on the probability of large cascades. 
This is typical cascading behavior. For example,
a simple branching process, with one initiating outage and $2896$ components, has  probability 0.05 of producing a cascade of size 50 or larger at $\lambda=0.92$. If this rate is reduced to $\lambda=0.79$, the probability of relatively small cascades is not changed much, but the probability of a cascade of 50 or more is reduced by 85\% and the probability of a cascade of size 100 or more is reduced by 96\%.

It is important to note that this method of finding critical upgrades differs substantially from conventional contingency ranking approaches. 
Whereas contingency ranking seek to identify components that would have substantial impact on the network if they occur as an initiating outage, our method seeks to find components that are important if they fail during a cascade. 
Regardless of this difference, one might conjecture that standard contingency ranking methods, such as the performance index approach in~\cite{Ejebe:1979}, might work equally as well, since they can be used to find components that substantially impact system loading levels.
To test whether this was indeed the case, we computed the performance index from~\cite{Ejebe:1979} for each transmission branch in the Polish test case and compared the result to $\alpha$ from (\ref{alpha_j}).
The two metrics showed only a weak correlation ($\rho=0.2$), and $\alpha$ was much more effective in selecting components for upgrades. 
Specifically, we used the performance index method to identify 10 branches for upgrade (doubling the flow capacity as before), and then simulated the 614 $n-2$ contingencies that produced at least 10 dependent transmission line outages in the re-modified case. 
While the re-modified case also produced smaller cascades relative to the original data, the influence graph method produced a much larger reduction in cascade sizes (see Table~\ref{PI_IG}).

\begin{table}
\caption{Average cascade sizes before and after upgrading 10 lines selected from either the 
performance index (PI) or influence graph (IG) methods}
\label{PI_IG}
\centering 
\begin{tabular}{cccc}
\hline 
 & Before & After, PI & After, IG \\
\hline 
Load shed & 5910 MW & 3682 MW & 1324 MW \\
Dependent outages & 55.8 & 42.5 & 14.1 \\
\hline 
\end{tabular}
\end{table}

\section{Discussion}

These results suggest that the influence graph approach can be used 
to simulate cascades that are statistically similar to those produced by an engineering simulator, 
and (more usefully) 
to quickly identify critical components and upgrades in a large power network.
However, it is important to note that the influence graph is a statistical model, and thus includes assumptions that one should be aware of.
Different assumptions are needed for the formulation, construction and usage of the model; 
the following paragraphs explain these assumptions.

The influence graph formulation at the beginning of Section II makes the standard branching process assumptions of 
independently generating the number of failures in the next generation from each of the failures in the current generation.
The offspring distributions $f$ are assumed to be Poisson 
 with parameters $\lambda_{i,m}$ that can vary with component $i$ and with generation $m$.
The model further assumes that the particular component failures that fail in the next generation can be modeled probabilistically as described in (3) for a single component failure in a generation, and that multiple failures in a generation propagate independently according to (3). These assumptions are also usefully discussed in \cite{LarremorePRE12}.
In addition, the model assumes that multiple outages in a generation each propagate independently to the next generation. 
(This is a standard and surprisingly successful assumption in applying branching processes.)
We acknowledge that outages can and do, in real power systems, mutually interact in producing outages in the next generation, 
but the experience with neglecting the branching process assumption and assuming independence
in power systems~\cite{RenCAS08,DobsonPStrips12} (and other subjects~\cite{LarremorePRE12}) is that good predictions of the total number of outages can nevertheless be made.
In power systems, Refs.~\cite{RenCAS08,DobsonPStrips12} validate the branching process assumption for the purpose 
 of predicting the total number of line outages using real line outages.
 Branching processes can match the distribution of number of simulated  line outages simulated by the OPA simulation in~\cite{DobsonRA10} and load shed simulated by  the OPA simulation~\cite{KimSJ12}.
 There is also a match for the distribution of load shed for the TRELSS simulation in one case of an industrial system of about 6250 buses~\cite{KimSJ12}.
Moreover, branching processes can analytically approximate CASCADE, a high-level probabilistic model of cascading~\cite{KimREL10} that 
explicitly models the additive effect of multiple line outages at each stage.

To estimate the parameters of the model (Section II.A), one needs to decide on a method for the division of the outages in each cascade into generations. If there are multiple outages in a generation, then the offspring of each outage are approximated by dividing by the number of outages.  An important assumption required for any statistical technique is that sufficient data was simulated for the estimation of parameters. The detail in representing the cascading, such as how much it depends on the cascade generation, must be traded off against this requirement for sufficient data.

In order to generate cascades with the influence graph in Section III, one needs an assumption about how to draw from $f$ and $g$. There are several possibilities here.  For example, one could draw to obtain precisely the number of outages indicated by the draw from $f$, but this would require adjusting the probabilities in $g$ to reflect the fact that some components are already outaged. To avoid this, we assume that the draw from $f$ indicates the number of times one should draw from $g$. Since no adjustments to were made to $g$ during this process, the result is that some components can be marked for failure multiple times.  Since real transmission components cannot outage multiple times without restoration, we count each failed component only once.

Finally, in order to perform the linear algebra in (16)-(23) and hence identify critical elements in Section IV, further approximations are needed.
Higher order probabilities are neglected to obtain the matrix multiplication in (15), and we need to allow the system to count the multiple failures of some components. In addition, in order to allow (19) we need to assume that $\mathbf{H}_{1+}$ does not change over the course of an arbitrarily long cascade. 

Our approach to estimate the parameters of $f$, 
which describe the propagation of outages, 
is based on standard Harris estimators from the theory of branching processes \cite{DobsonPStrips12}.
We use a rather straightforward approach to estimate the parameters of $g$, 
which describe the network structure and the interactions between outages. 
More sophisticated approaches may well significantly improve the performance of this estimation in future work, such as starting from Bayesian priors or adapting graph estimation algorithms from machine learning~\cite{Getoor07}.

There is a need for future work that evaluates each of these assumptions in detail, and works to identify ways to relax these assumptions without overcomplicating the model.

\section{Conclusions}
This paper presents a new approach to cascading failure risk analysis, in which we transform massive amounts of data from many cascading failures into an ``influence graph'' that describes the many ways that cascades propagate within that system.
We check that the (much simpler) influence graph conforms to the original data by comparing the distribution of cascading outage sizes from a cascading failure simulator to the distribution of cascade sizes generated from the influence graph. The two approaches produce remarkably similar cascade size distributions.
In addition, we derive a method with which the influence graph can be used to quickly identify component upgrades that can have a significant impact on cascading failure risk.

\section{Acknowledgement}
P.H. acknowledges helpful conversations with Cris Moore and Sid Redner at the Santa Fe Institute about these results.



\vspace{-10mm}

\begin{IEEEbiography}[{\includegraphics[width=1in]{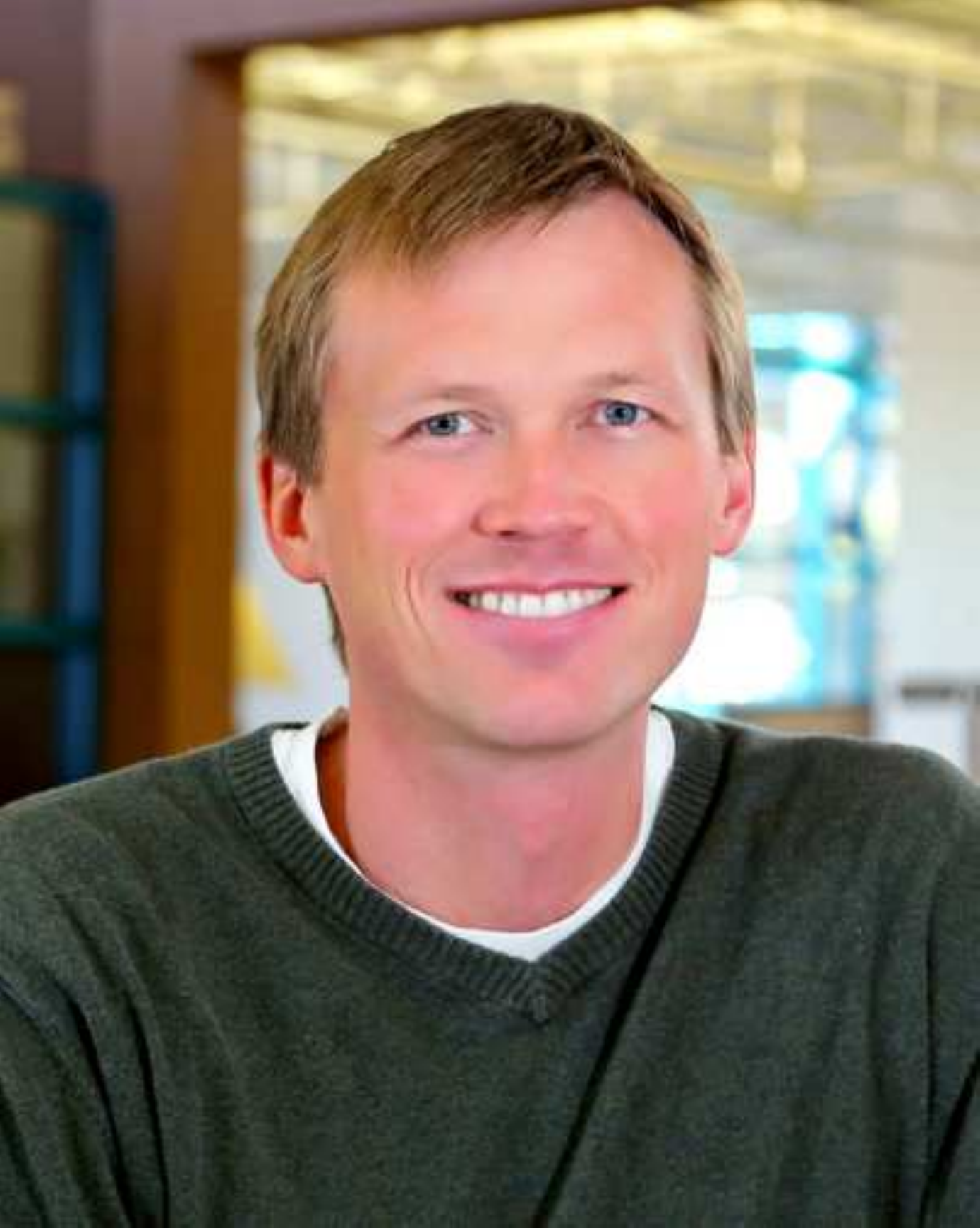}}]{Paul D.~H.~Hines} 
(S`96,M`07,SM`14) received the Ph.D. degree~in Engineering and Public Policy from Carnegie Mellon University in 2007 and 
M.S.~(2001) and B.S.~(1997) degrees in Electrical Engineering from the University of Washington and Seattle Pacific University, respectively. 

He is currently an Associate Professor, and the L.~Richard Fisher professor, 
in the Electrical Engineering Department at the University of Vermont. 
\end{IEEEbiography}

\vspace{-10mm}

\begin{IEEEbiography}[{\includegraphics[width=1in]{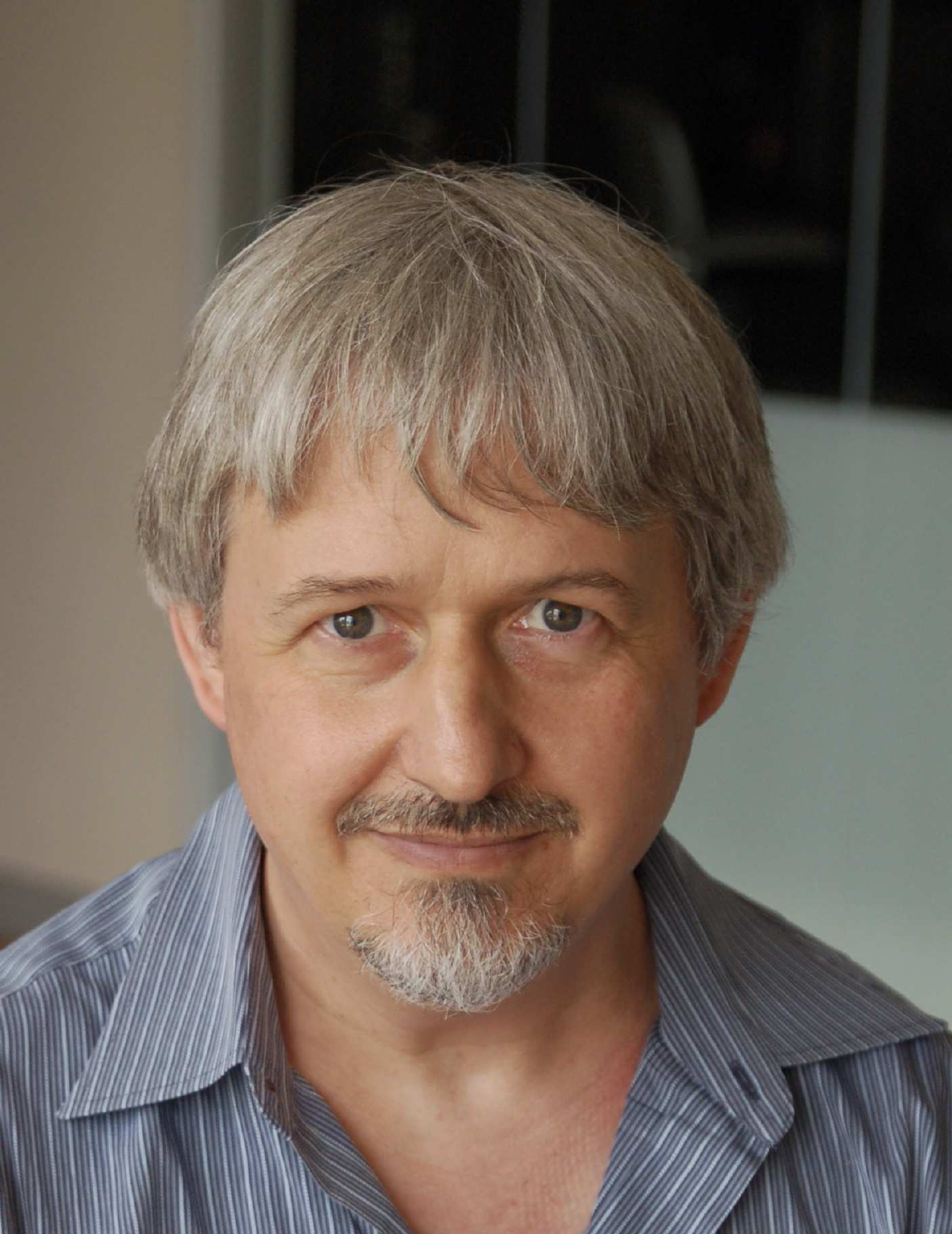}}]{Ian Dobson} 
(F'06) received the BA in Mathematics from Cambridge University and the PhD in Electrical Engineering from Cornell University. 

He is Sandbulte professor of electrical engineering at Iowa State University. 
His interests are cascading failure risk, complex systems, blackouts, electric power system stability, synchrophasors, and nonlinear dynamics.
\end{IEEEbiography}

\vspace{-10mm}

\begin{IEEEbiography}[{\includegraphics[width=1in]{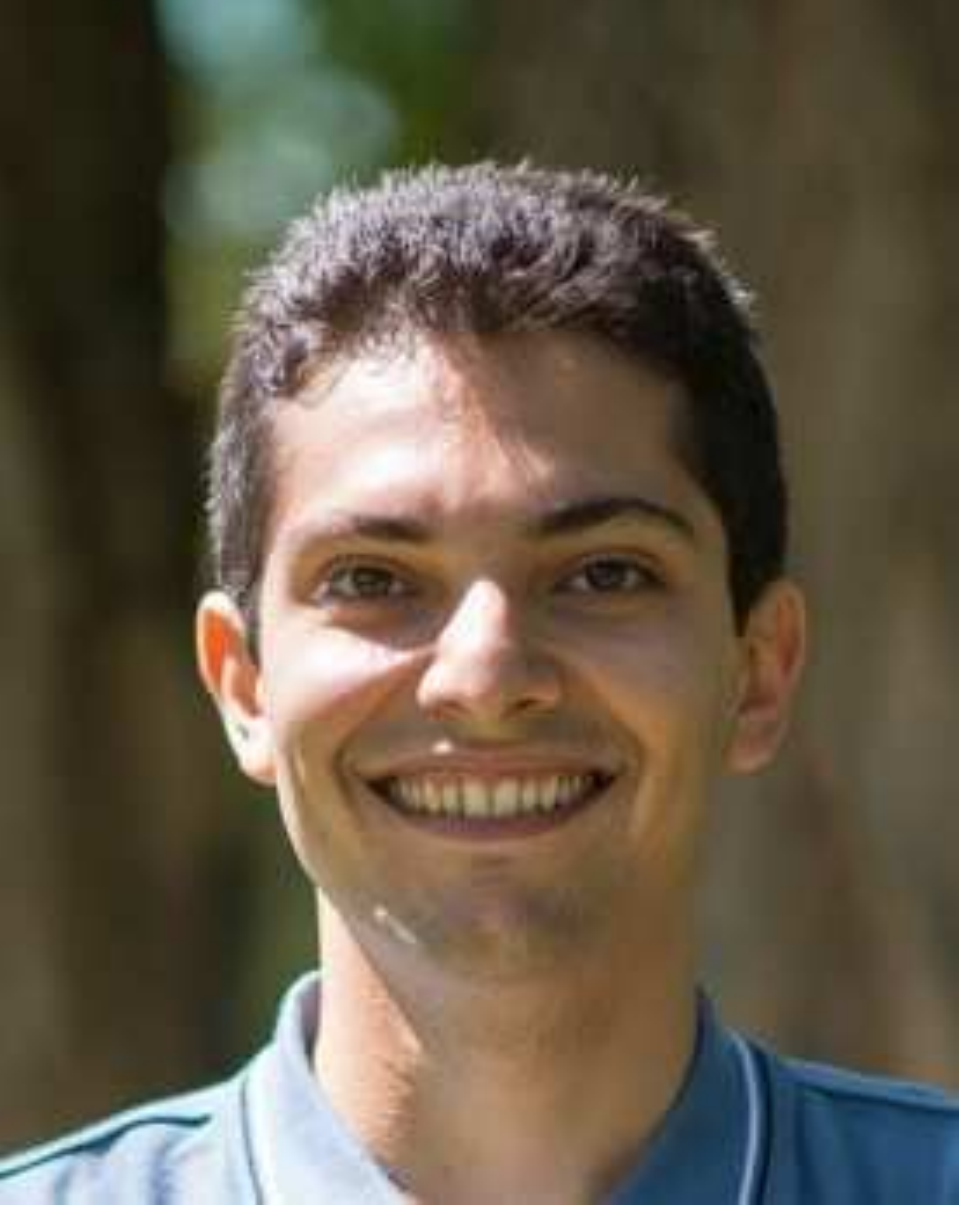}}]{Pooya Rezaei} (S`12) received the Ph.D. in Electrical Engineering from University of Vermont in 2015, and M.Sc. and B.Sc. degrees in Electrical Engineering from Sharif University of Technology (2010) and University of Tehran (2008), respectively. 
	
He is currently a Research Associate with the Digital Technology Center at the University of Minnesota. His research interests include smart grid, power system operation and control, and cascading failures in power networks. 
\end{IEEEbiography}

\end{document}